\documentclass[aps,prl,reprint,superscriptaddress,showpacs]{revtex4-1}

\usepackage{graphicx,amsmath,amssymb,ulem}
\usepackage{verbatim}
\usepackage{ragged2e}
\usepackage{color}

\begin{document}

\title{Multiphoton nonclassical light from clusters of single-photon emitters}

\author{Luo Qi}
\affiliation{Max Planck Institute for the Science of Light, Staudtstra\ss{}e 2, 91058 Erlangen, Germany}
\affiliation{University of Erlangen-N\"urnberg, Staudtstrasse 7/B2, 91058 Erlangen, Germany}
\author{Mathieu Manceau}
\affiliation{Max Planck Institute for the Science of Light, Staudtstra\ss{}e 2, 91058 Erlangen, Germany}
\author{Andrea Cavanna}
\affiliation{Max Planck Institute for the Science of Light, Staudtstra\ss{}e 2, 91058 Erlangen, Germany}
\affiliation{University of Erlangen-N\"urnberg, Staudtstrasse 7/B2, 91058 Erlangen, Germany}
\author{Fabian Gumpert}
\affiliation{University of Erlangen-N\"urnberg, Staudtstrasse 7/B2, 91058 Erlangen, Germany}
\author{Luigi Carbone}
\affiliation{CNR NANOTEC-Institute of Nanotechnology c/o Campus Ecotekne, University of Salento, via Monteroni, Lecce 73100, Italy}
\author{Massimo de Vittorio}
\affiliation{National Nanotechnology Laboratory - Istituto Nanoscienze CNR, Via Arnesano 16 - 73100 Lecce, Italy}
\author{Alberto Bramati}
\affiliation{Laboratoire  Kastler  Brossel,  UPMC-Sorbonne  Universités, CNRS,  ENS-PSL  Research  University,  Collège  de  France}
\author{Elisabeth Giacobino}
\affiliation{Laboratoire  Kastler  Brossel,  UPMC-Sorbonne  Universités, CNRS,  ENS-PSL  Research  University,  Collège  de  France}
\author{Lukas Lachman}
\affiliation{Department of Optics, Palacky University, 17 Listopadu 12, Olomouc 77146, Czech Republic}
\author{Radim Filip}
\affiliation{Department of Optics, Palacky University, 17 Listopadu 12, Olomouc 77146, Czech Republic}
\author{Maria V.~Chekhova}
\affiliation{Max Planck Institute for the Science of Light, Staudtstra\ss{}e 2, 91058 Erlangen, Germany}
\affiliation{University of Erlangen-N\"urnberg, Staudtstrasse 7/B2, 91058 Erlangen, Germany}

\begin{abstract}
We study nonclassical features of multiphoton light emitted by clusters of single-photon emitters. As signatures of nonclassicality, we use violation of inequalities for normalized correlation functions of different orders or the probabilities of multiphoton detection. In experiment,  for clusters of 2 to 14 colloidal CdSe/CdS dot-in-rods we observe nonclassical behavior, which gets even more pronounced for larger clusters.
\end{abstract}

\maketitle

\section{Introduction}
The development of quantum technologies requires advanced sources of nonclassical light. In particular, there is a challenge to go beyond single-photon states, two-photon states, and squeezed states, available at the moment. One option is to combine single-photon emitters into a group (a cluster) and obtain light with a limited number of photons~\cite{Shcherbina2014}. For such a cluster, the number of photons will not significantly exceed the number of emitters, which suggests applications in quantum key distribution~\cite{Scarani2009} and quantum metrology~\cite{Giovannetti2011}, where a limit on the maximal number of the photons is highly relevant. Recently, it has been predicted that even a large number of realistic single photon emitters  can produce nonclassical light~\cite{Shcherbina2014,Straka2018}. Importantly, this way of obtaining multiphoton quantum states does not require postselection or heralding, unlike methods based on nonlinear optical effects. 

The focus of this work is on the quantum properties of light emitted by clusters of colloidal quantum dots. By independently estimating the number of emitters in a cluster, we study the dependence of nonclassical features on this number. Using different nonclassicality criteria, we see that all of them are satisfied for clusters of up to 14 emitters, and some of them become even more pronounced as the number of emitters grows.

\section{Witnessing nonclassicality}
By definition, a nonclassical state is the one whose Glauber-Sudarshan (P) function is negative or singular~\cite{Leonhardt,DNK}.  However, because the P function is not directly measurable in experiment, various observable sufficient conditions of nonclassicality have been formulated. Putting aside loss-sensitive features, like squeezing or Wigner function negativity, here we will focus on those accessible through direct detection with a limited efficiency.

The simplest of them is antibunching, formulated in terms of the second-order normalized correlation function (CF) $g^{(2)}$,
\begin{equation}
g^{(2)}<1.
\label{eq:AB}
\end{equation}
Sometimes its analogs involving CFs of higher orders,
\begin{equation}
g^{(k)}<1, \,\,k>2.
\label{eq:higherAB}
\end{equation}
are referred to as `higher-order antibunching'~\cite{Stevens2013}. Third-order antibunching has been observed in Refs.~\cite{Stevens2013,Rundquist2014} for a single quantum dot coupled to a cavity.

More general inequalities indicating nonclassicality in terms of normalized CFs of different orders have been formulated by Klyshko~\cite{DNK}: 
\begin{equation}
g^{(k-1)}g^{(k+1)}<[g^{(k)}]^2.
\label{eq:nonclas}
\end{equation}

At $k=1$, Eq.~(\ref{eq:nonclas}) becomes the anti-bunching condition (\ref{eq:AB}) because $g^{(0)}=g^{(1)}=1$.
Accordingly, one can introduce a \textit{nonclassicality parameter of order} $k+1$, $\hbox{NP}(k+1)$, corresponding to the order of the highest CF it involves,
\begin{equation}
\hbox{NP}(k+1)\equiv g^{(k-1)}g^{(k+1)}-[g^{(k)}]^2.
\label{eq:N}
\end{equation}
Its negativity is an operational witness of nonclassicality~\cite{DNK}.

We would like to stress that conditions (\ref{eq:nonclas}) are stronger than `higher-order anti-bunching' (\ref{eq:higherAB}) in the sense that from all conditions (\ref{eq:nonclas}) up to $k+1$ satisfied, all conditions (\ref{eq:higherAB}) up to $k+1$ follow. For instance, $g^{(2)}<1$ in combination with $\hbox{NP}(3)<0$ leads to $g^{(3)}<1$.

Recently, alternative hierarchies of $k$th-order nonclassicality witnesses have been formulated, based on the `click' statistics of on-off detectors~\cite{Sperling2013,Filip2013,Lachman2015}. Compared to the conditions on the `click' statistics, an advantage of conditions (\ref{eq:nonclas}) is that they are formulated in terms of normalized CFs, which are invariant to optical losses or detection inefficiency~\cite{DNK}. On the other hand, witnesses of nonclassicality~\cite{Filip2013} or quantum non-Gaussianity~\cite{Straka2018} formulated in terms of multiphoton detection probabilities are more robust to noise than (\ref{eq:nonclas})~\cite{Moreva2017}.

In particular, the nonclassicality witnesses introduced in Ref.~\cite{Lachman2015} are $\theta^{(k)}<1$, where
\begin{equation}
\theta^{(k)}=\frac{P_{0^{\bigotimes k}}}{\prod_{i=1}^k P_{0[i]}},
\label{eq:theta}
\end{equation}
$P_{0^{\bigotimes k}}$ is the probability that neither of $k$ detectors `clicks', and $P_{0[i]}$ is the probability that the $i$th detector does not `click'. They have been applied to observe the nonclassical features of emission from ensembles of color centers in diamond~\cite{Moreva2017} and ions in a trap~\cite{Obsil2017}. However, the nonclassicality has been witnessed only for the case of $k=2$. In what follows, for the first time we test conditions (\ref{eq:nonclas}) and (\ref{eq:theta}) with $k=2,3,4$ for clusters of up to $14$ emitters. 

\section{Experiment}
\label{sec:experiment}
In our experiment we use colloidal semiconductor quantum dots~\cite{colloidal}. These emitters, although featuring a certain amount of blinking~\cite{blinking} and bleaching~\cite{bleaching}, and a non-negligible probability of two-photon emission, are very convenient due to their room-temperature operation and relatively simple production. The `dot-in-rod' (DR) modification~\cite{DR} is especially promising because of reduced blinking and a high degree of polarization~\cite{Pisanello,Manceau2014,Vezzoli2015}. Clusters are easily formed~\cite{Shcherbina2014,Israel2017} by dropping a DR solution onto a substrate and leaving the solvent to evaporate, the mean number of DRs in a cluster depending on the solution concentration. For this work we use CdSe/CdS DRs with $2.7$ nm core diameter, $22$ nm shell length and $4$ nm shell width. The DRs are dissolved in toluene with the concentration $10^{-14}$ mol/l and coated onto a fused silica cover slip thus obtaining a surface density of less than $0.1 \mu {\rm m}^{-2}$.

\begin{figure}[h!]
\begin{center}
\includegraphics[width=0.95\columnwidth]{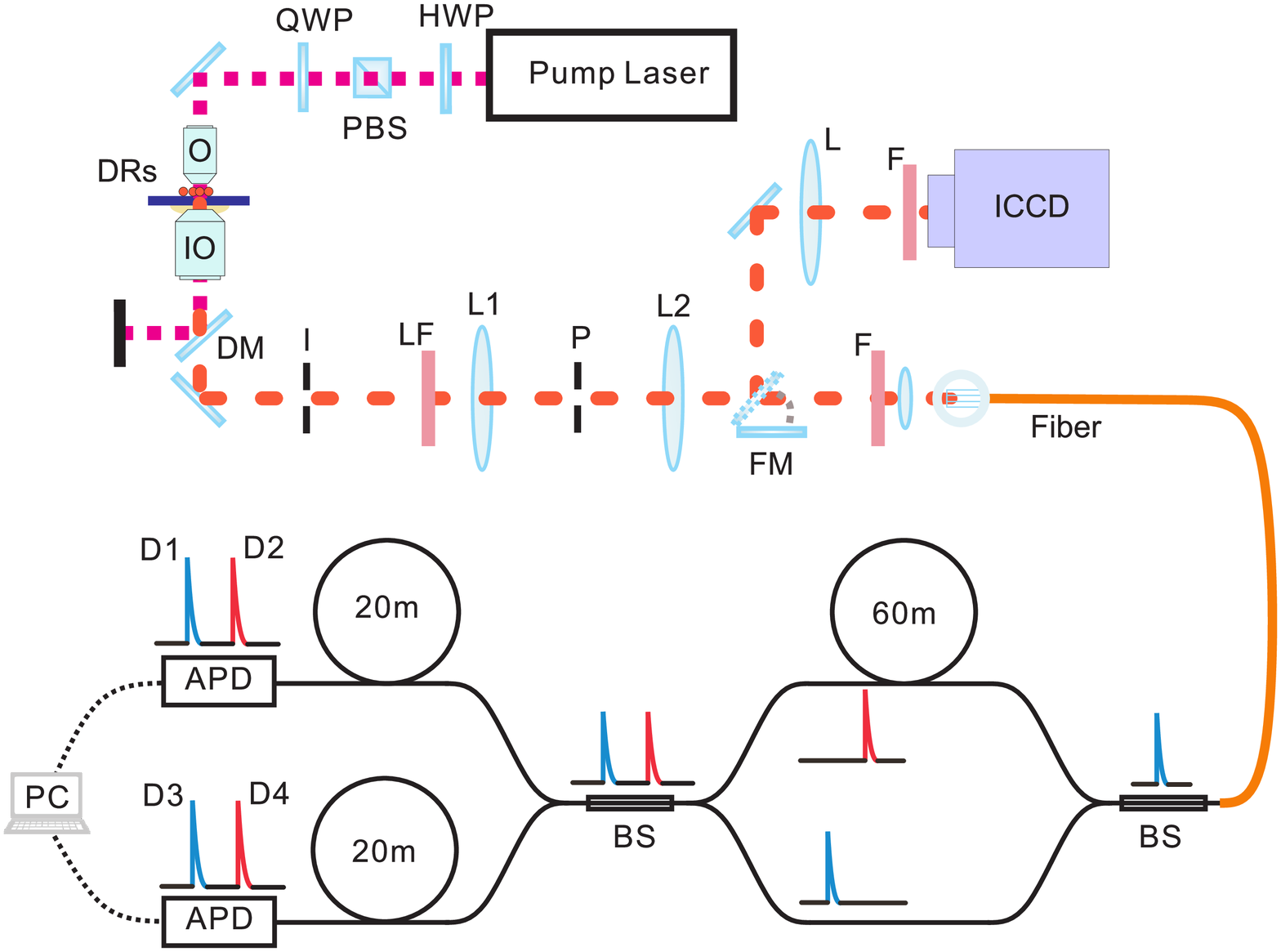}
\caption{The experimental setup. DR clusters are excited by the 3rd harmonic of a Nd:YAG laser through objective O. The emission is collected with an immersion objective IO and sent either into an ICCD camera, with a flip mirror FM, or into a fiber leading to a time-multiplexed HBT detection setup. For spatial filtering, a confocal scheme including lenses L1 and L2 and removable pinhole P is used. Frequency filtering is performed with long-pass filter LF and bandpass filters F.} \label{fig:setup}
\end{center}
\end{figure}
In the experimental setup (Fig.~\ref{fig:setup}), the excitation is with the pulsed third harmonic radiation of Nd:YAG laser with the wavelength $355$ nm, pulse duration $18$ ps and repetition rate $1$ kHz. The energy per pulse can be varied with the help of a half-wave plate HWP and a polarisation beamsplitter PBS; to reduce the probability of two-photon emission, it is chosen to be at $20\%$ of the saturation level. In this case, a single DR manifests strong anti-bunching with $g^{(2)}_1\le0.05$ and $g^{(3)}_1\le0.01$. A quarter-wave plate QWP transforms the polarization into circular, in order to provide the same excitation efficiency for all DRs regardless of their orientation. To uniformly excite many clusters of different sizes, the laser radiation is focused through an NA0.65 objective (O) placed on top of the sample, the illuminated area being $0.13$ mm large. A fused silica cover slip with the DRs on top is placed over an NA$1.3$ oil immersion objective (IO), so that more than $70\%$ of the emission is collected by the IO. Due to the use of thin fused silica cover slips the fluorescence noise is very low, leading to the signal-to-noise ratio higher than $3$ even for the smallest cluster under study.

\begin{figure}[h!]
\begin{center}
\includegraphics[width=0.9\columnwidth]{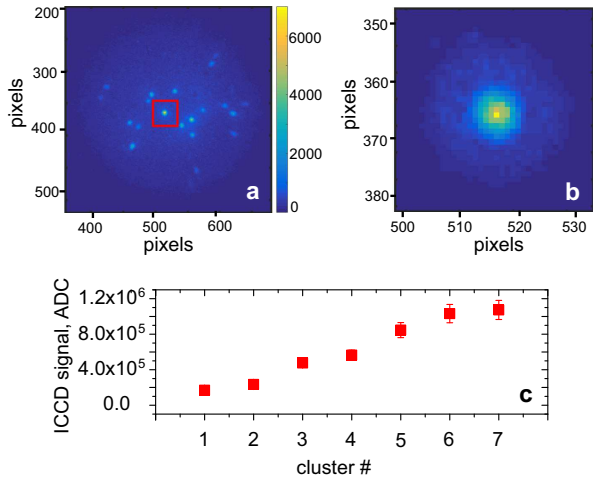}
\caption{Several DR clusters imaged by the ICCD camera (a), the zoomed image of one of the clusters (red frame in panel a), selected with the confocal pinhole (b), and the integral signal from each cluster, with the noise subtracted (c).} \label{fig:images}
\end{center}
\end{figure}

The DR emission is centered at $606$ nm and has a full width at half maximum (FWHM) of $40$ nm.  A dichroic mirror (DM) reflects the pump and transmits the DR emission into the registration part of the setup. A long-pass filter (LF) with the cutoff wavelength $570$ nm removes the remaining radiation of the pump. An intensified CCD (ICCD) camera (Princeton Instruments PI-MAX3:1024i) after a flip mirror (FM) is used to observe the image of several clusters and to choose ones containing different numbers of DRs. The image is formed by lens L with the focal distance $25$ cm. As an example, Fig.~\ref{fig:images}a shows the images of several clusters. The radiation from the chosen cluster is selected by an iris aperture I and sent, by removing the flip mirror M, through a multimode fiber into the Hanbury Brown - Twiss (HBT) setup using two avalanche photodiodes (APDs) and time multiplexing~\cite{Achilles2003,Fitch2003}. The time multiplexing scheme contains a $60$ m fiber loop (Fig.~\ref{fig:setup}), so that each of the APDs can receive photons in one of the two time slots, separated by $300$ ns. This scheme is then equivalent to a HBT setup with four detectors D1, D2, D3, D4, and enables the registration of up to fourfold coincidences and measurement of CFs of orders 2-4~\cite{Avenhaus2010}. To prevent cross-talk between the APDs, caused by flashes of light accompanying photon detections~\cite{Kurtsiefer}, $20$ m of fiber is inserted in front of each APD.

For eliminating background noise (caused by stray light and fluorescence of the substrate and other optical elements), filters (F) are placed in front of both the ICCD and the HBT setup: another long-pass filter (cutoff wavelength $570$ nm) and a bandpass filter (centered at $607$ nm, with $42$ nm FWHM). In addition, confocal microscopy filtering is arranged in front of the HBT setup by placing a $100\,\mu$m pinhole P between two confocal lenses (L1, L2) with focal lengths $7.5$ cm. The image of a single cluster with the pinhole present is shown in Fig.~\ref{fig:images}b.

The data are taken for seven clusters, having different brightness in the ICCD image (Fig.~\ref{fig:images}c). After subtracting the background noise, the integral output signal obtained for these clusters varies from $1.7\cdot10^5$ to $1.1\cdot10^6$ analog-to-digital conversion (ADC) units of the camera (Fig.~\ref{fig:images}c). As demonstrated in Ref.~\cite{Shcherbina2014}, this parameter can be used as an indicator of the number $m$ of DRs in a cluster. This, together with the measured value $g^{(2)}=0.413\pm0.009$ for the smallest cluster, allows us to conclude that the clusters contain from $m=1.7\pm0.2$ to $m=14\pm2$ DRs~\cite{fractional}. The mean number of photons per pulse detected from these clusters varies from $0.013$ to $0.20$. The low number of detected photons per pulse from a single DR (about $0.01$) is due to the low excitation rate ($10\%$) as well as the limited collection and detection efficiency.

For this reason, and because of the low rate of data acquisition ($1$ kHz), nearly no fourfold coincidences are acquired within several hours. Meanwhile, the number of detected two- and threefold coincidences suffices to measure the second- and third-order normalized CFs. This is done using the equation~\cite{Ivanova2006}
\begin{equation}
g^{(k)}=\frac{N_c^{(k)}}{N_1\dots N_k},
\label{eq:calc_g}
\end{equation}
where $N_c^{(k)}$ and $N_i$, $i=1\dots k$, are the mean numbers of $k$-fold coincidences and photon counts in the $i$-th detector, respectively, during a single pulse. Note that because of the low excitation and detection efficiency, $N_i\ll1$, hence the probabilities to have a `click' in the $i$th detector during a single pulse is $P_i\approx N_i\ll1$, justifying the validity of Eq.~(\ref{eq:calc_g})~\cite{Stevens2013}.

\begin{figure}[h!]
\begin{center}
\includegraphics[width=0.95\columnwidth]{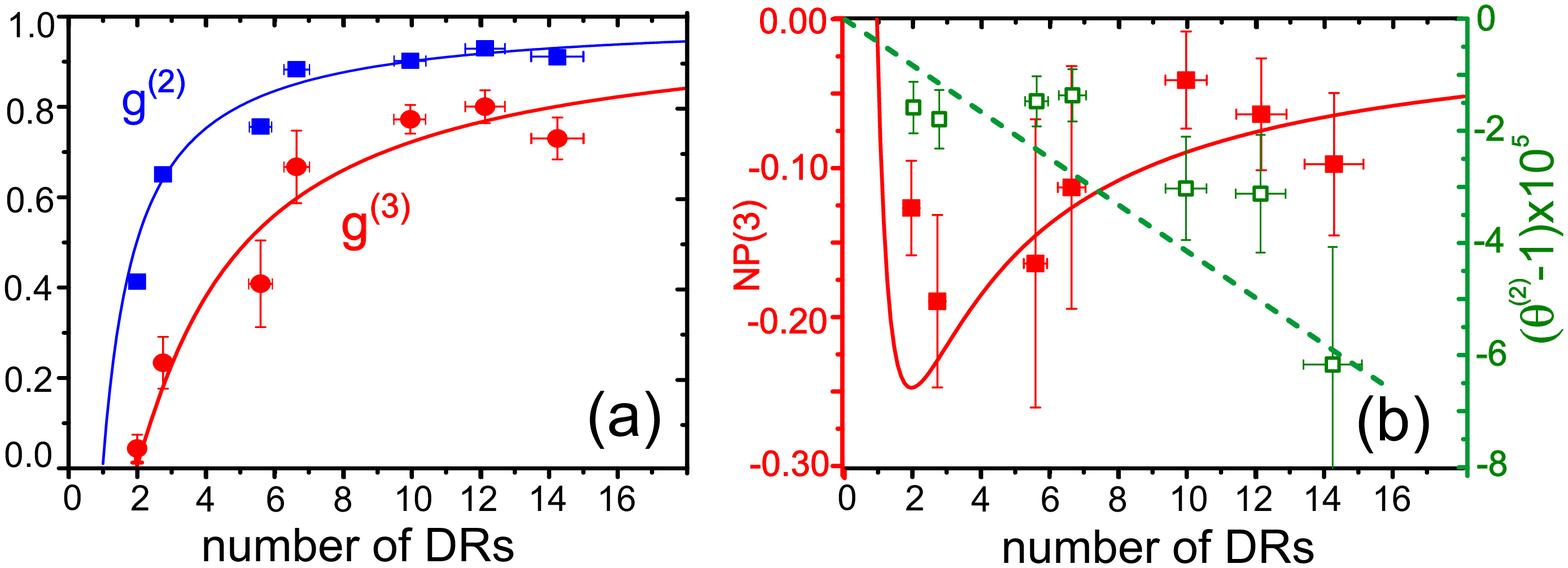}
\caption{The second- (blue squares) and third-order (red circles) CFs (a) and the nonclassicality parameters NP(3) (red circles) and $\theta^{(2)}-1$ (green squares) (b) measured for different clusters of DRs , versus the effective numbers $m$ of DRs in these clusters.} \label{fig:results}
\end{center}
\end{figure}
Fig.~\ref{fig:results}a shows the normalized CFs measured for different clusters and plotted as functions of their size, estimated from their brightness. The lines show the theoretical fits with the relations for the normalized CFs of a group of $m$ independent emitters. For the second-order CF of such a cluster, with the noise negligible~\cite{Shcherbina2014,Israel2017},
\begin{equation}
g^{(2)}_m=1+\frac{g^{(2)}_1-1}{m},
\label{eq:g2m}
\end{equation}
where $g^{(2)}_1$ is the normalized second-order CF of a single emitter. The third-order CF of such a cluster can be calculated to be~
\begin{equation}
g^{(3)}_m=1+\frac{g^{(3)}_1+3(m-1)g^{(2)}_1-3m+2}{m^2},
\label{eq:g3m}
\end{equation}
where $g^{(3)}_1$ is the third-order normalized CF of a single emitter.

For the fits in Fig.~\ref{fig:results}a, we used (\ref{eq:g2m}) and (\ref{eq:g3m}) with $g^{(2)}_1=0.01$ and $g^{(3)}_1=0$. The resulting curves are in a good agreement with the experimental points. All data points are below the unity, demonstrating antibunching of the second and third orders.

In Fig.~\ref{fig:results}b, red points show the third-order nonclassicality parameter $\hbox{NP}(3)$ for the same seven clusters, plotted versus their estimated size. One can see that it is negative for all clusters. This clearly indicates the nonclassicality; however, similarly to panel (a), the distance from the classical boundary decreases as the number of DRs in the cluster grows. The solid line is the theoretical dependence using expressions (\ref{eq:g2m},\ref{eq:g3m}).

The same panel shows the value of $\theta^{(2)}-1$ (\ref{eq:theta}), plotted with green empty squares. Here, unlike with the antibunching and $NP(3)$, the distance from the classical boundary increases with the increase in the size of the cluster. Indeed, if all emitters in a cluster have the same second-order CF, then~
\begin{equation}
\theta^{(2)}-1=Cm(g_1^{(2)}-1),
\label{eq:theta_exp}
\end{equation}
the parameter $C$ scaling quadratically with the detection efficiency. This dependence is shown in Fig.~\ref{fig:results}b with the dashed green line. Deviations of the experimental points from this line are due to the difficulty to control the coupling of the emission into the fiber; the uncertainty in $C$ reaches in this case 30\%. Note that if some of the emitters in a cluster have $g_1^{(2)}>1$, the dependence of $\theta^{(2)}$ on $m$ can be different~

Finally, for a chosen large cluster, containing $12\pm 1$ DRs, during about 30 hours of acquisition we obtained a set of data with up to four-fold coincidences. These data enabled the measurement of the normalized CFs $g^{(n)}$ and the parameters NP(n).

\begin{figure}[h!]
\begin{center}
\includegraphics[width=0.7\columnwidth]{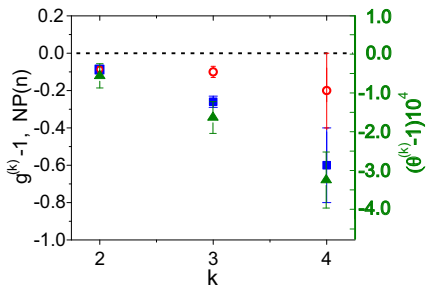}
\caption{The second- to fourth-order nonclassical features measured for a cluster of $12\pm 1$ DRs: $\hbox{NP}(k)$ (red empty circles), $g^{(k)}-1$ (blue filled squares) and $\theta^{(k)}-1$ (green filled triangles) as functions of $k$.} \label{fig:fourth}
\end{center}
\end{figure}

The results are shown in Fig.~\ref{fig:fourth}. All three normalized CFs of orders $2-4$ (blue filled squares) are well below the unity, demonstrating, for the first time to the best of our knowledge, up to the fourth-order antibunching.
Meanwhile, the Klyshko nonclassicality parameters $\hbox{NP}(k)$ (red empty circles) show negativity exceeding the measurement error only for $k=2,3$. Verification of nonclassicality for $k=4$ requires more experimental data.
At the same time, condition $\theta^{(k)}-1<1$ is satisfied for all $k=2,3,4$, its violation growing with $k$.  The reason is that in the limit of low detection probabilities $P_i$, expressions (\ref{eq:theta}) for $k=3,4$ become~
\begin{equation}
\theta^{(k)}-1\approx C\frac{k(k-1)}{2}m(g_1^{(2)}-1),
\label{eq:theta_exp}
\end{equation}
the deviation from the classical boundary increases both with the number of emitters $m$ and with the order $k$.
Note that the latter tendency is the same for all nonclassicality parameters: the larger the order $k$, the larger the deviation from the classical boundary. At the same time, conditions $\theta^{(k)}<1$ require less experimental data for verification than other conditions. 

\section{Conclusion}
In conclusion,  we have tested higher-order nonclassicality for clusters of 2-14 DRs. In addition to antibunching and its third- and fourth-order analogs, we have observed third-order Klyshko's nonclassicality, which has been shown to be a stronger condition. The low rate of detected single-photon emission ($1\%$ per excitation pulse) does not allow us to test higher-order antibunching or to witness fourth-order Klyshko nonclassicality. At the same time, for the nonclassicality parameters (\ref{eq:theta}) it was possible to overcome the classical boundary up to the fourth order, the deviation growing both with the order and with the size of the cluster. The observed nonclassical features are important for quantum key distribution and quantum metrology, where nonclassical light with limited number of photons is required.

\section*{Funding Information}
L.L. and R.F. acknowledge the financial support of the Czech Science Foundation (GB14-36681G). L.L. acknowledges the financial support of Palacký University (IGA-PrF-2017-008).

\section*{Acknowledgments}

The authors thank M.~Sondermann, V.~Salakhutdinov, and M.~Grassl for helpful discussions.



\begin{thebibliography}{99}
\bibitem{fractional}{A fractional number of DRs means that different DRs can have different emission efficiencies.}
\end{thebibliography}
\end{document}